\documentclass[twocolumn,pra,showpacs,superscriptaddress]{revtex4}

\usepackage{graphicx}
\usepackage{dcolumn}
\usepackage{bm}
\usepackage{amssymb}
\usepackage{amsmath}

\begin{document}

\title{Valley-dependent gauge fields for ultracold atoms in square optical superlattices}

\author{Dan-Wei Zhang}
\affiliation{Laboratory of Quantum Engineering and Quantum
Materials, SPTE, South China Normal University, Guangzhou 510006,
China}

\author{Chuan-Jia Shan}
\affiliation{College of Physics and Electronic Science, Hubei Normal
University, Huangshi 435002, China} \affiliation{Laboratory of
Quantum Engineering and Quantum Materials, SPTE, South China Normal
University, Guangzhou 510006, China}

\author{Feng Mei}
\affiliation{National Laboratory of Solid State Microstructures and
School of Physics, Nanjing University, Nanjing 210093, China}

\author{Mou Yang}
\affiliation{Laboratory of Quantum Engineering and Quantum
Materials, SPTE, South China Normal University, Guangzhou 510006,
China}

\author{Rui-Qiang Wang}
\affiliation{Laboratory of Quantum Engineering and Quantum
Materials, SPTE, South China Normal University, Guangzhou 510006,
China}

\author{Shi-Liang Zhu}
\email{slzhunju@163.com} \affiliation{National Laboratory of Solid
State Microstructures and School of Physics, Nanjing University,
Nanjing 210093, China}

\date{\today}

\begin{abstract}

We propose an experimental scheme to realize the valley-dependent
gauge fields for ultracold fermionic atoms trapped in a
state-dependent square optical lattice. Our scheme relies on two
sets of Raman laser beams to engineer the hopping between adjacent
sites populated by two-component fermionic atoms. One set of Raman
beams are used to realize a staggered $\pi$-flux lattice, where
low energy atoms near two inequivalent Dirac points should be
described by the Dirac equation for spin-1/2 particles. Another
set of laser beams with proper Rabi frequencies are added to
further modulate the atomic hopping parameters. The hopping
modulation will give rise to effective gauge potentials with
opposite signs near the two valleys, mimicking the interesting
strain-induced pseudo-gauge fields in graphene. The proposed
valley-dependent gauge fields are tunable and provide a new route
to realize quantum valley Hall effects and atomic valleytronics.

\end{abstract}

\pacs{67.85.-d, 03.75.Lm, 03.65.Pm}

\maketitle

The low-energy effective theory of graphene describes relativistic
Dirac fermions near the two inequivalent corners of the Brillouin
zone, termed valleys \cite{Neto}. Valley index plays an important
role in the extraordinary electronic properties of graphene.
Valley-dependent gauge fields, usually called pseudo-gauge fields to
distinguish from the real valley-independent electromagnetic field
in unstrained graphene, have recently been studied extensively both
theoretically \cite{Guinea2006,Pereira,Guinea2009} and
experimentally \cite{Levy,Guinea2010}. It has been show that such
gauge fields can be realized by modulating the electronic hopping
with strains in a two-dimensional (2D) honeycomb lattice
\cite{Levy,Guinea2010}. These findings open up an exciting area of
mechanically engineering band structure of graphene \cite{Pereira},
as well as realizing some exotic phenomena absent in other
solid-state materials, such as new types of quantum Hall related
effects \cite{Guinea2009,Ghaemi}.

On the other hand, a growing class of Dirac materials with
synthetic honeycomb structure have recently been proposed and
explored \cite{Polini}, such as trapped cold atoms in optical
lattices (OL) \cite{Zhu2007,Esslinger}, confined photons in
photonic crystals \cite{Rechtsman,Bellec}, and molecular graphene
\cite{Gomes}. Interestingly, the pseudo-magnetic fields and
related Landau levels have been experimentally demonstrated in
photonic graphene \cite{Rechtsman} and molecular graphene
\cite{Gomes} by designing a spatial texture of hopping parameters.
In addition, the creation and manipulation of Dirac points with a
Fermi gas in a honeycomb OL have been also reported recently
\cite{Esslinger}. A promising extension in this cold atom system
is to simulate the tunable valley-dependent gauge fields and
realize the related novel effects. For this purpose, a practical
way is to modulate the atomic hopping parameters in a honeycomb OL
by using the synthetic gauge potentials \cite{Dalibard} or the
laser-assisted tunneling (LAT) \cite{Jaksch,Gerbier}, following
the schemes proposed in Refs. \cite{Zhu2008,Alba2011,Alba2013}.
However, the LAT technique has not yet been demonstrated in
honeycomb OLs, but in square optical (super) lattices
\cite{Bloch2011,Bloch2013,Miyake}. Therefore, a natural question
is whether one can simulate the valley-dependent gauge fields
within current experimental technique in a square OL.

In this Brief Report, we propose a feasible scheme to realize the
valley-dependent gauge fields for ultracold fermionic atoms trapped
in a square optical superlattice. In our scheme, a state-dependent
square OL populated by two-component atoms is considered and this
lattice has a checkerboard configuration, which allows for
engineering LAT in the two spatial directions. As the first step to
simulate the valley-dependent gauge fields, two Raman laser beams
are employed to create a staggered $\pi$-flux lattice, which results
in an effective relativistic Hamiltonian near the two inequivalent
Dirac points. The second step is to further modulate the hopping
amplitudes in the previous $\pi$-flux lattice by using another two
or three Raman beams with proper Rabi frequencies. If the hopping
modulation is smooth over the lattice spacing scale, it will give
rise to effective gauge potentials with opposite signs near the two
valleys, mimicking the interesting pseudo-gauge fields in strained
graphene
\cite{Guinea2006,Pereira,Guinea2009,Levy,Guinea2010,Ghaemi}. These
synthetic gauge fields can be controlled by carefully designing the
Rabi frequencies of the Raman laser beams in the second step. In
addition, we briefly present some potential applications with these
gauge fields, including the quantum valley Hall effect (QVHE) and
atomic valleytronics. Although there has been a great deal of
theoretical and experimental studies in producing artificial gauge
fields for neutral atoms \cite{Dalibard}, none of them couple with
the valley degree of freedom. So our proposal can enlarge the
community of gauge fields in cold atom systems and provide a pathway
towards realizing atomic valley-based devices.

\begin{figure}[tbph]
\vspace{0.5cm} \includegraphics[width=8.2cm]{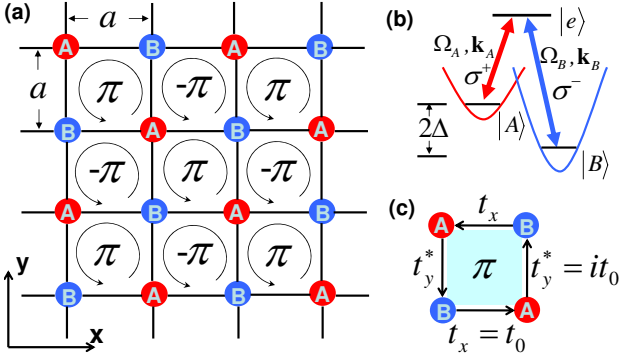} 
\caption{(Color online) (a) Schematic two-dimensional
state-dependent OL with a checkerboard configuration and a
staggered $\pi$-flux. (b) Sketch of laser-assisted tunneling
between two nearest-neighbor lattice cites. The atoms with
internal ground states $|A\rangle $ and $|B\rangle $ are
respectively trapped in sublattices A and B, with a tunable
on-site energy imbalance. They are coherently coupled to the
excited state $ |e\rangle $ through Raman laser beams with Rabi
frequencies $\Omega _{A}e^{i \mathbf{k}_{A}\cdot \mathbf{r}}$ and
$\Omega _{B}e^{i\mathbf{k}_{B}\cdot \mathbf{r}}$. (c) The hopping
configuration of each plaquette for simulating a staggered
$\pi$-flux lattice \protect\cite{Gerbier}, with the hopping
parameters along the $\hat{x}$ ($\hat{y}$) axis as $t_{x}=t_{0}$
($t_{y}=-it_{0}$). For a particle hopping anticlockwise around a
plaquette, the phase factor picked up along the path
$(i,j)_{B}\rightarrow (i+1,j)_{A}\rightarrow
(i+1,j+1)_{B}\rightarrow (i,j+1)_{A}\rightarrow (i,j)_{B}$ is
$e^{i\protect\pi}$. For the adjacent plaquette, the phase factor
is $e^{-i\protect\pi}$.}
\end{figure}

Let us start by considering a 2D noninteracting two-component
fermionic gas in a state-dependent square OL with a checkerboard
structure, as shown in Fig. 1(a). Such state-dependent OLs have
been experimentally created by superposing two linearly polarized
laser beams with a relative polarized angle, where the separation
and potential depth of the two sublattices (A and B) can be well
controlled by the angle and the laser intensity \cite{Bloch2003}.
The fermionic atoms are commonly chosen as $^{6}$Li or $^{40}$K in
current experiments. For $^{6}$Li atoms, the hyperfine levels for
the two-component states can be
$|A\rangle =|2^{2}S_{1/2},\frac{3}{2},-\frac{1}{2}\rangle $ and
$|B\rangle =|2^{2}S_{1/2},\frac{3}{2},\frac{3}{2}\rangle $,
respectively. For $^{40}$K atoms, the hyperfine levels can be
$|A\rangle =|4^{2}S_{1/2},\frac{7}{2},-\frac{1}{2}\rangle $ and
$|B\rangle =|4^{2}S_{1/2},\frac{7}{2},\frac{3}{2}\rangle $. In this
OL, the atoms must alter their internal states in order to tunnel
between two nearest-neighbor lattice sites. This can be achieved by
the so-called LAT method
\cite{Jaksch,Gerbier,Bloch2011,Bloch2013,Miyake}. Two Raman laser
beams with Rabi frequencies $\Omega _{A}e^{i\mathbf{k}_{A}\cdot
\mathbf{r}}$ and $\Omega _{B}e^{i\mathbf{k}_{B}\cdot \mathbf{r}}$
are applied to couple the states $|A\rangle $ and $|B\rangle $ via
an immediate excited state $|e\rangle $, as shown in Fig. 1(b). The
excited states for $^{6}$Li and $^{40}$K atoms are chosen as
$|2^{2}P_{1/2},\frac{1}{2},\frac{1}{2}\rangle $ and
$|4^{2}P_{1/2},\frac{9}{2},\frac{1}{2}\rangle$, respectively. So the
polarizations of the two Raman transition lasers are $\sigma ^{+}$
and $\sigma ^{-}$.

Through adjusting the Raman laser parameters appropriately, a
magnetic $\pi$-flux lattice \cite{Jaksch,Gerbier,Bloch2011}
illustrated in Fig. 1(a) can be simulated and we will show this in
the following. The tight-binding Hamiltonian of the lattice system
takes the form
\begin{equation}
H_{0}=-\sum_{\langle i,j\rangle }\left( t_{ij}\hat{a}_{i}^{\dag }\hat{b}_{j}+%
\text{ H.c.}\right) +\Delta \sum_{i}\left( \hat{a}_{i}^{\dag }\hat{a}_{i}-%
\hat{b}_{i}^{\dag }\hat{b}_{i}\right) ,  \label{Ham}
\end{equation}
where $\hat{a}_{i}^{\dag }$ ($\hat{b}_{i}^{\dag }$) is the creation
operator for the internal state $|A\rangle $ ($|B\rangle $) at
lattice site $i$ belonging to the sublattice A (B), $\langle
i,j\rangle $ denotes the nearest-neighbor hopping with the hopping
parameter $t_{ij}=-\int w_{A}^{\ast
}(\mathbf{r}-\mathbf{r}_{i})\Omega _{\text{eff}}w_{B}(\mathbf{r}-
\mathbf{r}_{j})d^{2}\mathbf{r}$ and $\Omega _{\text{eff}}=\Omega
_{A}^{\ast }\Omega _{B}e^{i(\mathbf{k}_{B}-\mathbf{k}_{A})\cdot
\mathbf{r}}$, the spatial coordinate $\mathbf{r}=\{x,y\}$ and
$w_{A,B}$ being the Wannier functions of the lowest Bloch band, and
$2\Delta$ is the tunable on-site energy imbalance between the two
sublattices. For proper laser beams, we can assume
$t_{ij}=t_{0}e^{iA_{ij}}$, where $t_{0}>0$ is the hopping magnitude
controlled by $\Omega _{A,B}$ and the overlap integral between the
Wannier functions associated with each sublattice, and $A_{ij}$ is
the phase induced by the wave vectors $\mathbf{k}_{A,B}$ in the LAT
process \cite{Jaksch,Gerbier}. For the staggered $\pi $-flux lattice
shown in Fig. 1(a), we have $\sum_{\circlearrowleft }A_{ij}=\pm \pi
$ for each plaquette. There are many approaches to generate the
lattice with phase $A_{ij}$. A practical method with the hopping
parameters $t_{x}=t_{0}$ and  $t_{y}=-it_{0}$ from sublattice B to
sublattice A, as shown in Fig. 1(c), can be found in Ref.
\cite{Gerbier}. Similar LAT schemes have been experimentally
realized with bosonic atoms \cite{Bloch2011,Bloch2013,Miyake}.

By using the Fourier transformation $\hat{a}_{j}=\frac{1}{\sqrt{N}}\sum_{%
\mathbf{k}}e^{i\mathbf{k}\cdot \mathbf{r}_{j}}\hat{a}_{\mathbf{k}}$ and $%
\hat{b}_{j}=\frac{1}{\sqrt{N}}\sum_{\mathbf{k}}e^{i\mathbf{k}\cdot \mathbf{r}%
_{j}}\hat{b}_{\mathbf{k}}$ on the Hamiltonian in Eq. (1) with
$t_{x}=t_{0}$ and $t_{y}=-it_{0}$, we can obtain the corresponding
Hamiltonian in the momentum space as
\begin{equation}
\label{Hamk} {H}_{{\bf k}0}=\sum_{\bf k}\left(\hat{a}^{\dag}_{\bf
k}, \hat{b}^{\dag}_{\bf k}\right)\left(
                                          \begin{array}{cc}
                                            \Delta & f_{\bf k} \\
                                            f_{\bf k}^\ast & -\Delta \\
                                          \end{array}
                                        \right)
\left(
                                        \begin{array}{c}
                                          \hat{a}_{\bf
k} \\
                                          \hat{b}_{\bf
k} \\
                                        \end{array}
                                      \right),
\end{equation}
where $f_{\mathbf{k}}=-2t_{0}[\cos (k_{x}a)-i\cos (k_{y}a)]$ with
$a$ being the lattice spacing. Thus the energy spectrum is given
by $E(\mathbf{k})=\pm \sqrt{|f_{\mathbf{k}}|^{2}+\Delta ^{2}}$,
which exhibits two inequivalent Dirac points at $K_{\pm }=\pm
\frac{\pi }{2a}(1,1)$ with an energy gap $2\Delta$. By
substitution of $\mathbf{k}\rightarrow K_{\pm }+\mathbf{q}$, the
dynamics around the Dirac points $K_{\pm }$ (i.e.,
$|\mathbf{q}|a\ll 1$) is then governed by the effective Dirac
Hamiltonian \cite{Neto,Zhu2007}
%
\begin{equation}
H_{\eta }=\eta \hbar v_{F}\left( \sigma _{x}q_{x}+\sigma
_{y}q_{y}\right) +\Delta \sigma _{z},  \label{HamEff}
\end{equation}
where $\eta =\pm $ represent different valleys $K_{\pm }$,
$v_{F}=2t_{0}a/\hbar$ is the effective Fermi velocity, and $\sigma
_{i}$ are the Pauli matrices with $i=\{x,y,z\}$. This low-energy
effective Dirac Hamiltonian is similar to the unstrained nature
graphene and the sublattice degree of freedom here  plays the role
of the spin degree of freedom.

\begin{figure}[tbph]
\vspace{0.5cm} \includegraphics[width=8.2cm]{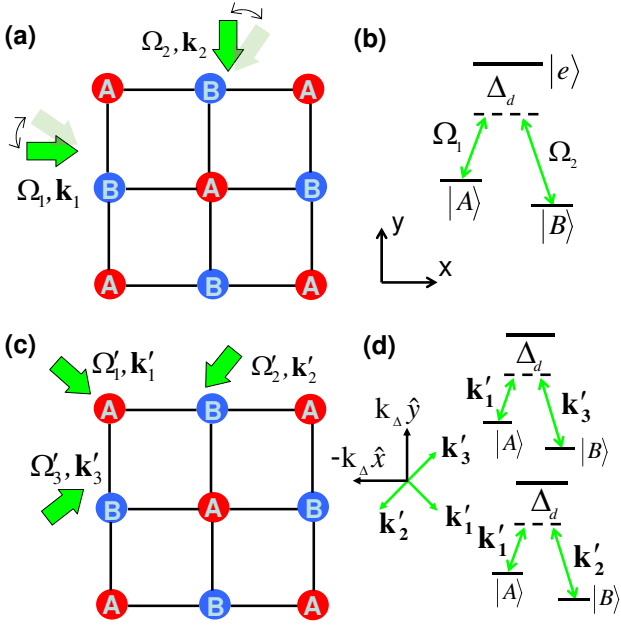}
\caption{(Color online) Two schemes (a, b) and (c, d) for
generating tunable valley-dependent gauge potentials. (a) Two
additional Raman beams for modulating the atomic hopping with
adjustable wave vectors. (b) Sketch of the additional laser
coupling between the states $|A\rangle$ and $|B\rangle$ with large
detuning $\Delta_d$ from the state $|e\rangle$. (c) Three
additional Raman beams for modulating the atomic hopping with
fixed wave vectors. (d) Sketch of the additional coupling with the
three lasers and the effective
wave numbers $\mathbf{k}^{\prime }_2-\mathbf{k}^{\prime }_1=-k_{\Delta}\hat{x%
}$ and $\mathbf{k}^{\prime }_3-\mathbf{k}^{\prime }_1=k_{\Delta}\hat{y}$.}%
\end{figure}

Note that similar Dirac Hamiltonian has been proposed in
state-independent square OLs with other methods for light-induced
gauge potentials \cite{Lim}, especially the related quantum
anomalous Hall phase was also investigated in a square
checkerboard lattice \cite{Goldman}. Although the Dirac-like
equation and gauge fields have been studied in
Refs.\cite{Lim,Goldman}, none of them explored the
valley-dependent gauge fields for cold atomic systems. These gauge
fields have been firstly proposed and then experimentally realized
in natural and artificial graphene \cite{Levy,Rechtsman,Gomes}.
However, it is unclear how to create them on other lattice
geometries instead of the honeycomb lattice \cite{Guinea2010}. In
the following, we will propose two approaches to realize the
valley-dependent gauge fields for cold atoms trapped in the square
optical superlattices by modulating the hopping parameters.

To realize a valley-dependent gauge fields based on LAT method, we
assume that the hopping amplitudes along $\hat{x}$ and $\hat{y}$
axis in the previous $\pi$-flux system are further modulated by
$\delta t_{x}$ and $\delta t_{y}$, respectively. The perturbation
Hamiltonian for such modulation is written as
\begin{equation}
\delta H=-\sum_{\langle i,j\rangle _{x}}\delta
t_{x}\hat{a}_{i}^{\dag }\hat{b }_{j}-\sum_{\langle i,j\rangle
_{y}}\delta t_{y}\hat{a}_{i}^{\dag }\hat{b} _{j}+\text{H.c.}.
\label{Ham}
\end{equation}
This modulation is done by using two Raman laser beams with
detuning $\Delta _{d}$, which give rise to the additional
couplings being shown in Figs. 2(a) and (b). The Rabi frequencies
of the two lasers are  $\Omega
_{1}(\mathbf{r})e^{i(\mathbf{k}_{1}\cdot \mathbf{r}+\phi _{1})}$
and $\Omega _{2}(\mathbf{r})e^{i(\mathbf{k }_{2}\cdot
\mathbf{r}+\phi _{2})}$, where $\mathbf{k}_{l}$ ($\phi _{l}$) with
$l=\{1,2\}$ are the wave vectors (phases) of the laser beams.
Defining $\theta =\phi _{2}-\phi _{1}$ and
$\mathbf{k}_{2}-\mathbf{k}_{1}=k_{x}\hat{x}+k_{y}\hat{y}$, we have
$\delta t_{x}=\int w_{A}^{\ast }(x_{j},y_{j})\delta \Omega
e^{i(k_{x}x+\theta )}w_{B}(x_{j\pm 1},y_{j})dxdy$ and $\delta
t_{y}=\int w_{A}^{\ast }(x_{j},y_{j})\delta \Omega
e^{i(k_{y}y+\theta )}w_{B}(x_{j},y_{j,j\pm 1})dxdy$ for the LAT
\cite{Jaksch,Gerbier}, where $ \delta \Omega \equiv \Omega
_{1}\Omega _{2}/\Delta _{d}$ and $w_{\alpha}(x_{j},y_{j})$ is the
Wannier function on the $\alpha$ (=A,B) lattice with lattice
position $(x_{j},y_{j})$. Generally, the modulation of the hopping
parameters can be rewritten as
\begin{equation}
\delta t_{x}=\delta t_{0}^{x}e^{i\varphi _{x}},~~\delta t_{y}=\delta
t_{0}^{y}e^{i\varphi _{y}}.
\end{equation}
Here $\delta t_{0}^{x,y}$ are the magnitudes determined by the
overlap integral with respect to $\delta \Omega ({\mathbf{r}})$ and
the Wannier functions, and $\varphi _{x,y}$ are the phases
determined by $k_{x,y}$ and $\theta $
\cite{Jaksch,Gerbier,Bloch2011,Bloch2013,Miyake}, thus the hopping
modulation can be easily tuned by adjusting the parameters of the
Raman laser beams. To preserve the staggered $\pi$-flux lattice and
keep the previous Dirac Hamiltonian around the Dirac points, the
additional flux in each plaquette has to be tuned as $2(\varphi
_{x}+\varphi _{y})=2N\pi $ with $N=\{0,\pm 1,\pm 2,\cdots \}$. This
can be achieved in experiments by appropriately adjusting the
parameters $k_{x}$, $k_{y}$, and $\theta$
\cite{Bloch2011,Bloch2013,Miyake}. For example, one may set
$k_{x}=k_{y}=\pi /(2a) $ and $\theta =0$, which yields $\varphi
_{x}=\varphi _{y}=\pi /2$ and satisfies the previous flux condition.
We also require $|\delta t_{0}^{x}|,|\delta t_{0}^{y}|\ll t_{0}$ as
a perturbation term, which can be satisfied by setting $|\delta
\Omega |\ll |\Omega _{\text{eff}}|$. For example, one may tune the
ratio between $\delta t_{0}^{x}$ ($\delta t_{0}^{y} $) and $t_{0}$
through adjusting the detuning $\Delta _{d}$.

With the previous Fourier transformation, the perturbation
Hamiltonian in the momentum space is given by
\begin{equation}
\delta H_{\mathbf{k}}=\sum_{\mathbf{k}}{\delta f}_{\mathbf{k}}\hat{a}_{%
\mathbf{k}}^{\dag }\hat{b}_{\mathbf{k}}+\text{H.c.},
\end{equation}%
where ${\delta f}_{\mathbf{k}}=-2[\delta t_{0}^{x}\cos
(k_{x}a+\varphi _{x})-i\delta t_{0}^{y}\cos (k_{y}a+\varphi
_{y})]$. If the modulation of the atomic hopping $\delta
t_{0}^{x}$ and $\delta t_{0}^{y}$ is smooth over the lattice
spacing scale, there is no Fourier component with $K_{+}-K_{-}$,
and thus the two Dirac valleys are decoupled by the perturbation
\cite{Neto,Guinea2010}. Within this smooth perturbation, the total
Hamiltonian in the momentum space $H_{\mathbf{k}}=H_{\mathbf{k}
0}+\delta H_{\mathbf{k}}$ at the vicinity of the two Dirac points
can be replaced by a low energy effective spinor Hamiltonian
\cite{Guinea2010}
\begin{equation}
\hat{\mathcal{H}}_{\text{eff}}^{\eta }=\eta
v_{F}\boldsymbol{\sigma }\cdot \left(
\boldsymbol{\hat{p}}+\boldsymbol{\mathcal{A}}\right) +\Delta
\sigma _{z},  \label{HamDirac}
\end{equation}
where $\boldsymbol{\sigma }=(\sigma _{x},\sigma _{y})$,
$\boldsymbol{\hat{p}}=(\hat{p}_{x},\hat{p}_{y})$ is the momentum
operator, and
$\boldsymbol{\mathcal{A}}=(\mathcal{A}_{x},\mathcal{A}_{y})$ with
$\mathcal{A}_{x}=\frac{2 }{\hbar v_{F}}\text{Re}[\delta
f_{\mathbf{k=K_{+}}}]$ and $\mathcal{A}_{y}= \frac{2}{\hbar
v_{F}}\text{Im}[\delta f_{\mathbf{k=K_{+}}}]$. Here we expand the
perturbation up to the first order of $\delta t_{0}^{x}/t_{0}$ and
$\delta t_{0}^{y}/t_{0}$. We obtain $\mathcal{A}_{x}$ and
$\mathcal{A}_{y}$ as
\begin{equation}
\begin{array}{ll}
\label{Gaugefield} \mathcal{A}_{x}({\bf r}) = \frac{2}{\hbar
v_F}\cos(\varphi_x+\frac{\pi}{2})\delta t_0^x({\bf r}),\\
\\
\mathcal{A}_{y}({\bf r}) = \frac{2}{\hbar
v_F}\cos(\varphi_y+\frac{\pi}{2})\delta t_0^y({\bf r}).
\end{array}
\end{equation}

The Hamiltonian (\ref{HamDirac}) describes the dynamics of Dirac
fermions in the presence of valley-dependent gauge potentials
\cite{Guinea2010}. It is obvious that $\mathcal{A}_{x}$ and
$\mathcal{A}_{y}$ play the role of gauge potentials with opposite
sign $\eta $ in different Dirac valleys. Here
$\boldsymbol{\mathcal{A}}(\mathbf{r})$ is tunable through adjusting
the parameters of Raman laser beams. For example, we can change the
spatial configuration of the Rabi frequencies $\Omega _{l}$
($l=1,2$) to control the position dependence of $\delta t_{0}^{x}$
and $\delta t_{0}^{y}$, and adjust the wave vectors $\mathbf{k}_{l}$
to tune $\varphi _{x}$ and $\varphi _{y}$, as demonstrated in
experiments \cite{Bloch2011,Bloch2013,Miyake} and shown in Fig.
2(a). In this way, however, we may be unable to independently tune
the spatial distributions of $\mathcal{A}_{x}$ and $\mathcal{A}_{y}$
since both of them depend on $\Omega _{1}(\mathbf{r)}$ and $\Omega
_{2}(\mathbf{r)} $.

To enhance the tunability of the valley-dependent gauge potentials
in this system, we can use three Raman beams \cite{Chan} instead of
two, as shown in Fig. 2(c). The Rabi frequencies of three lasers are
denoted by $\Omega
_{1}^{\prime}(\mathbf{r})e^{i\mathbf{k}_{1}^{\prime }\cdot
\mathbf{r}}$, $\Omega _{2}^{\prime
}(\mathbf{r})e^{i(\mathbf{k}_{2}^{\prime }\cdot \mathbf{r }+\theta
^{\prime })}$, and $\Omega _{3}^{\prime }(\mathbf{r})e^{i(\mathbf{k}
_{3}^{\prime }\cdot \mathbf{r}+\theta ^{\prime })}$, where
$\mathbf{k} _{m}^{\prime }$ the wave vectors with $m=\{1,2,3\}$ and
$\theta ^{\prime }$ the relative phase. We assume the directions of
the lasers are fixed as those in Fig. 2(d), with
$\mathbf{k}_{2}^{\prime }-\mathbf{k}_{1}^{\prime }=-k_{\Delta
}\hat{x}$ and $\mathbf{k}_{3}^{\prime }-\mathbf{k}_{1}^{\prime
}=k_{\Delta }\hat{y}$. Then the effective hopping-modulation
parameters along $\hat{x}$ and $\hat{y}$ axis ($\delta t_{x}$ and
$\delta t_{y}$) are replaced by $\delta t_{x}^{\prime }=\int
w_{A}^{\ast }(x_{j},y_{j})\delta \Omega _{x}e^{i(-k_{\Delta
}x+\theta ^{\prime })}w_{B}(x_{j\pm 1},y_{j})dxdy$ and $\delta
t_{y}^{\prime }=\int w_{A}^{\ast }(x_{j},y_{j})\delta \Omega
_{y}e^{i(k_{\Delta }y+\theta ^{\prime })}w_{B}(x_{j},y_{j,j\pm
1})dxdy$, with $\delta \Omega _{x}\equiv \Omega _{1}^{\prime }\Omega
_{2}^{\prime }/\Delta _{d}$ and $\delta \Omega _{y}\equiv \Omega
_{1}^{\prime }\Omega _{3}^{\prime }/\Delta _{d}$. In this case, we
can independently tune the spatial distributions of
$\mathcal{A}_{x}$ and $\mathcal{A}_{y}$ through adjusting $\Omega
_{2}^{\prime }$ and $\Omega _{3}^{\prime }$, respectively.

When the hopping modulations $\delta t_{0}^{x}$ and (or) $\delta
t_{0}^{y}$ are time dependent, the gauge potential in Eq.
(\ref{Gaugefield}) also becomes time dependent, as the one in nature
graphene under time-dependent strains \cite{Vaezi,Oppen}. Therefore
in general cases, we have $\boldsymbol{\mathcal{A}}(\mathbf{r},t)$,
which is associated with a valley-dependent effective
electromagnetic field $\{\mathbf{E}_{\eta },\mathbf{B}_{\eta }\}$
given by
\begin{equation}
\begin{array}{ll}
\label{EMfield}\mathbf{E}_{\eta }=-\eta \partial
\boldsymbol{\mathcal{A}}/{
\partial t},~~\mathbf{B}_{\eta }=\eta \nabla \times \boldsymbol{\mathcal{A}}.
\end{array}
\end{equation}
The time-dependent modulations in this system can be easily
realized by varying the Rabi frequencies (i.e., their laser
intensities) of the Raman beams or the detuning with time.
Interestingly, a time-dependent but valley-independent vector
potential associated with an effective electric field for neutral
atoms was created by tuning the detuning in the laser-atom
coupling with time \cite{Spielman2011}. In contrast, we have
proposed a feasible scheme to realize the tunable valley-dependent
gauge fields (including electric field) in a square optical
superlattice. We note that our scheme can be extended to simulate
the non-Abelian $SU(2)$ valley-dependent gauge fields by
introducing density waves or double layers with proper additional
optical potentials, similar to the proposals in molecular graphene
\cite{Ryu} and bilayer graphene \cite{Jose}.

We now present some potential applications with the tunable
valley-dependent gauge fields in this system. Firstly, we consider
QVHE with valley-Landau levels (VLLs) \cite{Guinea2009,Levy} which
requires a uniform valley-dependent magnetic field
$\mathbf{B}_{\eta }=\eta B_{0}\hat{z}$ ($ B_{0}>0$), corresponding
to the Landau gauge $\boldsymbol{\mathcal{A}} =(-B_{0}y,0)$.
Realization of this gauge potential requires carefully designing
the laser configurations \cite{note} and would be challenge as the
case in graphene \cite{Guinea2009}. The eigenstates of
$\hat{\mathcal{H}}_{ \text{eff}}^{\eta }$ with $\Delta =0$ then
fall into the quantized VLLs at energies
$E_{n}=\text{sgn}(n)\sqrt{2\hbar v_{F}^{2}B_{0}|n|}$ \cite{Neto}.
Because $\mathbf{B}_{\eta }$ has opposite signs for carriers in
valleys $K_{+}$ and $K_{-}$, the chiral edges states protected by
the bulk gap $ |E_{n+1}-E_{n}|$ at the boundary are
counterpropagating with different valleys. This is in contrast to
the copropagating ones for a real magnetic field, and is
reminiscent of topological insulators  and therefore, is called
QVHE characterized by zero charge Chern number and a nonzero
valley Chern number \cite{Zhang}. Interestingly, the $n=0$ VLL
wave functions in both valleys are localized entirely on the
sublattice B \cite{Ghaemi}. In our system, atoms in the sublattice
B have the internal state $|B\rangle $, which is convenient for
spin-resolved observation. Secondly, this system with tunable
valley-dependent electromagnetic fields provides an ideal platform
to explore atomic valleytronics \cite{Rycerz}. When the system is
metallic, the valley-dependent electric field can be used to drive
valley currents and further to design atomic valley filters
\cite{Rycerz,Ji}. We can also simulate valley Hall effects by the
valley-dependent magnetic field. In this case, the pseudo-magnetic
field can be nonuniform and even zero as
$\boldsymbol{\mathcal{A}}$ is a nonzero constant, which is much
easer to be achieved by selecting the laser configurations
\cite{note}. Note that atomic spin Hall effects \cite{Zhu2006}
have been observed in a very recent experiment
\cite{Spielman2013}.
Similarly, by subjecting the system to a valley-dependent electric
field, we may produce an atomic topological edge current related to
the valley degree of freedom \cite{Vaezi}.

Finally, we briefly discuss  the feasibility of our proposal with
the practical experimental parameters. Let us consider $^{40}$K
atoms with the typical lattice spacing $a\simeq 400$ nm and
lattice depth $V_{0}\simeq 22E_{r}$ \cite{Pachos2013}, where
$E_{r}$ is the recoil energy. With the typical choice,
$E_{r}/\hbar \simeq 8$ kHz, numerical simulations in Ref.
\cite{Pachos2013} indicate that the band gap between the two
lowest Bloch bands $ \Delta E_{\text{gap}}\simeq 8E_{r}$ and the
natural (next-nearest-neighbor) hopping within sublattices
$t_{N}\lesssim 10^{-3}E_{r}$. The nearest-neighbor hopping
$|t_{ij}|\equiv t_{0}$ is proportional to the effective Raman
intensity $|\Omega _{\text{eff}}|$, and $t_{0}\gtrsim \hbar
|\Omega _{\text{eff}}|\beta$ with the overlap integral of Wannier
functions between neighbor lattice sites $\beta \simeq 10^{-2}$.
So, a feasible value $ |\Omega _{\text{eff}}|\sim 10E_{r}/\hbar $
would not pump the atoms into the higher Bloch bands. For typical
hopping perturbation $\delta t_{0}^{x,y}\sim 0.1t_{0}\gtrsim
10t_{N}$, the natural hopping terms can be neglected safely. The
spontaneous emission and the associated atomic heating are also
negligible within several seconds in current experiments
\cite{Bloch2011,Bloch2013,Miyake}.

In summary, we have proposed an experimental scheme to simulate the
tunable valley-dependent gauge fields with ultracold fermionic atoms
in a square optical superlattice using the LAT method. Our scheme
provides a pathway to explore the quantum valley Hall effects and
atomic valleytronics. In view of the fact that the LAT technique has
been demonstrated in similar OLs in recent experiments
\cite{Bloch2011,Bloch2013,Miyake}, it is anticipated that the
present proposal will be tested in an experiment in the near future.

This work was supported by the NSFC (Grant No. 11125417 and No.
11274124), the SKPBR (Grant No. 2011CB922104), NCET (Grant No.
10-0090), the PCSIRT, and the SRFGS of SCNU.


\begin{thebibliography}{10}
\bibitem{note} The magnitude of the hopping perturbation is approximately
proportional to the effecive intensity of the additional Raman
beams. Therefore, to induce the uniform pseudo-magnetic field, one
may require $|\delta \Omega(x,y)|\propto y$ along $\hat{x}$ axis.
For the running-wave Raman lasers, the synthtic gauge potential is
position-independent and then $\mathbf{B}_{\eta}=0$; for the
standing-wave or Gaussian-type lasers, $\mathbf{B}_{\eta}$ is
position-dependent but nonuniform.
\end{thebibliography}
\end{document}